\documentclass[aps,showpacs,preprint,preprintnumber,nofootinbib,amsmath,amssymb,ascmac,12pt]{revtex4}
\usepackage{bm}
\usepackage[dvipdfm]{graphicx}
\begin{document}
\title{
Acceleration of colliding shells around a black hole
\\
{\it
--
Validity of the test particle approximation in the Banados-Silk-West process
--
}
}
\author{
${}^{1}$Masashi Kimura\footnote{E-mail:mkimura@sci.osaka-cu.ac.jp},
${}^{1}$Ken-ichi Nakao\footnote{E-mail:knakao@sci.osaka-cu.ac.jp}
and
${}^{2}$Hideyuki Tagoshi\footnote{E-mail:tagoshi@vega.ess.sci.osaka-u.ac.jp}
}
\affiliation{
${}^{1}$Department of Mathematics and Physics,
Graduate School of Science, Osaka City University,
3-3-138 Sugimoto, Sumiyoshi, Osaka 558-8585, Japan
\\
${}^{2}$Department of Earth and Space Science,
Graduate School of Science, Osaka University, Osaka 560-0043, Japan.
\vspace{1cm}
}
\begin{abstract}
Recently, Banados, Silk and West (BSW) showed that
the total energy of two colliding test particles 
has no upper limit in their center of mass frame 
in the neighborhood of an extreme Kerr black hole, 
even if these particles were at rest at infinity in the 
infinite past. We call this mechanism 
the BSW mechanism or BSW process. 
The large energy of such particles would generate strong gravity, although this has not 
been taken into account in the BSW analysis.
A similar mechanism is seen
in the collision of two spherical test shells 
in the neighborhood of an extreme Reissner-Nordstr\"om black hole.
In this paper, 
in order to draw some implications concerning the effects of gravity generated by 
colliding particles in the BSW process, we study a collision of 
two spherical dust shells, since their gravity can be exactly treated. 
We show that the energy of two colliding shells in the center of mass frame 
{\it observable from infinity} has an upper limit due to their own gravity. 
Our result suggests that an upper limit also exists for the total energy of colliding particles 
in the center of mass frame in the observable domain 
in the BSW process due the gravity of the particles. 
\end{abstract}
\preprint{OCU-PHYS 335}
\preprint{AP-GR 80}
\preprint{OU-TAP 312}
\pacs{04.70.Bw}
\date{\today}
\maketitle

\section{Introduction}\label{sec:intro}

Recently, Banados, Silk and West (BSW) showed that
two test particles can collide with arbitrarily high energy in the 
center of mass frame near an extremal Kerr black hole, even though 
these particles were at rest at infinity in the infinite past~\cite{Banados:2009pr}. 
We call this mechanism the BSW mechanism or the BSW process,
and several aspects of this mechanism have been reported
in Refs.~\cite{Berti:2009bk,Jacobson:2009zg,Grib:2010dz,Lake:2010bq,Grib:2010bs,Wei:2010vca,Zaslavskii:2010aw,Wei:2010gq,Mao:2010di,Harada:2010yv,Grib:2010xj,Banados:2010kn}.
If this mechanism was really workable, it might be possible to probe Planck-scale physics by 
observing the neighborhood of an extreme or almost extreme Kerr black hole. 
However, it is not yet clear whether particles can really be accelerated 
with sufficient efficiency to produce collisions with Planckian energies. 
To answer this question, it is necessary to consider, among other things, the effect of particle size,
the effect of gravitational radiation on the trajectories of the particles,
and the effect of the gravity generated by the particles themselves at the event horizon.
In this paper, we focus on the third effect.

The BSW mechanism is also interesting from a purely relativistic point of view.
The result obtained in Ref.~\cite{Banados:2009pr} seems to imply that
the energy of the colliding particles in the center of mass frame 
can be as large as the mass of the Kerr black hole in the background spacetime, 
and hence the gravity generated by the particles is so strong that another black hole forms. 
Such intense gravity can not be described using a linear perturbation approximation of the Kerr spacetime. 
Thus, even if the energy of the particles is initially small enough that the gravity 
due to these particles is well described by the linear perturbation approximation, 
the gravity generated by the energy of these particles might finally become 
too strong to invoke linear perturbation analysis. If this inference is true, 
the Kerr black hole is unstable against minor  perturbations induced by dropping small particles into the black hole under finely tuned initial conditions. 
This conclusion is paradoxical, because it is well known that
a Kerr black hole is stable against small perturbations.
Therefore, it is important to investigate how large the energy of colliding particles 
in the center of mass frame can be, by taking into account the 
gravity due to the particles.

However, it is difficult to treat the effects of gravity generated by 
particles around a Kerr black hole. 
One of the reasons for this difficulty is 
the fact that Kerr spacetime is not very symmetric. 
Hence, in the present paper, we consider charged particles or charged 
spherical shells around
a spherically symmetric  charged black hole known as a 
Reissner-Nordstr\"om black hole, and then show that a process 
similar to the BSW process is also possible in this system. 
Next, in order to draw some implications concerning the effects of the gravity 
generated by colliding objects in this process, we study a collision of 
spherical dust shells whose gravity is exactly treatable 
by the Israel formalism~\cite{Israel:1966rt}.

The organization of this paper is as follows.
In Sec.~\ref{sec:Kerr}, we briefly review the BSW 
mechanism. In Sec.~\ref{sec:RN}, we 
study a collision between a charged test particle and a neutral 
test particle around an extreme 
Reissner-Nordstr\"om black hole, in order to show that a BSW-like process is also possible in this spacetime. 
We study a spherical charged dust shell 
in~Sec.~\ref{sec:RNshell}, 
and a collision between the charged dust shell and a neutral dust shell 
in Sec.~\ref{sec:RNshellcollision}.
The final section is devoted to the discussion.

In this paper, we adopt geometric units for  which Newton's gravitational 
constant $G$ and the speed of light $c$ are unity,
and 
we adopt an abstract index notation in which all Latin indices except for $t$, $r$ and $i$ 
indicate a type of a tensor, whereas all Greek indices except for $\phi$ represent components with respect to 
the coordinate basis. The exceptional indices $t$, $r$, $\phi$ 
respectively denote the components of time, radial and azimuthal coordinates 
in the spherical polar coordinate system, and $i$ specifies the $i$-th particle, shell or region. 
The signature of the metric is diag$[-,+,+,+]$. 

\section{Particle Collisions around Extreme Kerr Black Holes}\label{sec:Kerr}

In this section, we briefly review the results described in Ref.~\cite{Banados:2009pr}.

We consider a collision of two test particles with an identical inertial mass $m$. 
We denote the 4-velocities of these particles by $u_{(i)}^a$ ($i=1,2$). 
Then, the total 4-momentum of these particles at the collision event is given by 
$p^a=m(u_{(1)}^a+u_{(2)}^a)$, and their total energy in the 
center of mass frame, which hereafter will be called simply the CM energy, is given by
\begin{eqnarray}
E_{\rm cm}= \sqrt{-g_{ab}p^a p^b}
=\sqrt{2} m \sqrt{1 - g_{ab}u_{(1)}^a u_{(2)}^b},
\label{cm}
\end{eqnarray}
where $g_{ab}$ is the metric tensor.

Let us consider the CM energy at a collision event of two particles in a 
Kerr spacetime whose metric is given by
\begin{eqnarray}
ds^2 &=& -\left(1-\frac{2 Mr}{\rho^2}\right)dt^2
-\frac{4 aM r \sin^2\theta}{\rho^2}dtd\phi
+\frac{\rho^2}{\Delta}dr^2
\notag\\ & &
+
\rho^2 d\theta^2
+
\left(r^2 + a^2 +
\frac{2 a^2 Mr \sin^2\theta}{\rho^2}
\right)\sin^2 \theta d\phi^2,
\label{Kerr}
\end{eqnarray}
where $M$ and $a$ are the mass parameter and the Kerr parameter which represents the 
angular momentum of this system, respectively, and 
\begin{eqnarray}
\Delta &=& r^2 + a^2 - 2 M r,
\\
\rho^2 &=& r^2 + a^2 \cos \theta. 
\end{eqnarray}
In the case of $M^2>a^2$, the equation $\Delta=0$ has two real roots 
$r=r_\pm:=M\pm\sqrt{M^2-a^2}$: $r=r_+$ corresponds to the black hole (BH) 
horizon, whereas $r=r_-$ corresponds to the Cauchy horizon. In the case 
of $M^2=a^2$, these two horizons degenerate into one horizon $r=r_+=r_-=M$. 
In the case of $M^2<a^2$, there is no root of $\Delta=0$, and hence this case is 
naked singular. In this section, we assume $M\geq|a|$.

For simplicity, we focus on two particles whose orbits are confined to the 
equatorial plane $\theta = \pi/2$. By integrating the geodesic equations and using 
the normalization condition of the 4-velocity 
$g_{ab}u_{(i)}^a u_{(i)}^b=-1$ ($i=1,2$), we have 
\begin{eqnarray}
u_{(i)}^r &=& \pm \frac{1}{r^2} \sqrt{
T_i^2 - \Delta [
r^2 + (\ell_i - a {\cal E}_i)^2
]
},
\label{geor}
\\
u_{(i)}^\phi &=& -\frac{1}{r^2}
\left[
(a{\cal E}_i - \ell_i) - \frac{a T_i}{\Delta}
\right],
\label{geophi}
\\
u_{(i)}^t &=& -\frac{1}{r^2}
\left[
a (a{\cal E}_i - \ell_i) - \frac{(r^2 + a^2) T_i}{\Delta}
\right],
\label{geot}
\end{eqnarray}
where ${\cal E}_i$, $\ell_i$ are constants of integration, 
and $T_i = {\cal E}_i(r^2 + a^2) -\ell_i a$. Note that ${\cal E}_i$ and $\ell_i$ 
correspond to the specific energy and angular 
momentum of the $i$-th particle.

We assume that each particle is marginally bound : ${\cal E}_i=1$. In this case, $u_{(i)}^r$ and 
$u_{(i)}^\phi$ vanish in the limit of $r\rightarrow\infty$, or in other words, 
these particles were at rest at infinity in the infinite past. 
By substituting Eqs.~(\ref{geor})--(\ref{geot}) into Eq.~(\ref{cm}), 
we obtain the square of the CM energy at the collision event as
\begin{eqnarray}
E_{\rm cm}^2 &=& \frac{2 m^2}{r(r^2 - 2Mr +a^2)}
\bigg(
2a^2 (r+M)
-
2 Ma (\ell_1 + \ell_2)
-
\ell_2 \ell_1
(r -2M)
+
2 
(r -M)r^2
\notag\\ &&\qquad
-
\sqrt{2M(a-\ell_1)^2 - \ell_1^2 r + 2M r^2}
\sqrt{2M(a-\ell_2)^2 - \ell_2^2 r + 2M r^2}
\bigg).
\end{eqnarray}
In the extreme case $a = M$, the CM energy at the degenerate horizon $r=r_\pm=M$ 
is given in the simple form
\begin{eqnarray}
E_{\rm cm}(r \to r_+) = \sqrt{2}m \sqrt{
\frac{\ell_2 - 2M}{\ell_1 - 2M}
+
\frac{\ell_1 - 2M}{\ell_2 - 2M}
}.
\end{eqnarray}
We see from the above equation that 
if either $\ell_1$ or $\ell_2$ is equal to $2M$ and the 
other is not, the CM energy at the collision event does not have an upper limit. 
\footnote{
More generally, the case of the Kerr Newmann black holes is discussed in \cite{Wei:2010vca}.
}

This result can be understood as follows. 
We denote the world line of the 1st particle by $x^\mu=x^\mu(\tau)$, where $\tau$ 
is its proper time. 
Then, in the extreme case $a = M$, the square of (\ref{geor}) gives 
\begin{eqnarray}
\left(\frac{dr}{d\tau}\right)^2 - 
\frac{2M(r - M)^2 }{r^3}
=0,
\label{eq1}
\end{eqnarray}
where we have taken ${\cal E}_1 = 1$ and $\ell_1 = 2M$. 
We can see from this equation 
that if this particle is falling toward the black hole, 
its asymptote is the degenerate horizon $r = M$.
Since the only possible causal line 
on the future horizon is outward null, 
this particle asymptotically becomes outward null, even though it 
is falling toward the black hole\footnote{One might expect that an unstable circular
orbit for a massive particle 
is possible at the horizon $r= M$ from Eq. (\ref{eq1}).  
However, we can show that this is not true. The detailed discussions are given in \cite{Harada:2010yv}.}. 
Hence, the closer the relative velocity between this 1st particle and a 2nd particle 
with $\ell_2\neq 2M$ approaches to the speed of light, 
the closer the collision event approaches the BH horizon. 
As a result, the collision of these particles can lead to an indefinitely 
large CM energy near the horizon. 
Here, note that the 1st particle cannot reach the horizon within a finite 
time span, and hence the CM energy never diverges.

\section{Particle Collision around Extreme Reissner-Nordstr\"om Black Holes}\label{sec:RN}

Since the large CM energy of the particles 
produces strong gravity, we must take this into account when evaluating how large the CM energy can become in the BSW mechanism. 
However, it is difficult to treat the effects of gravity due to the particles, partly because Kerr spacetime is not very symmetric.

However, a mechanism similar to the BSW mechanism has been reported for 
Reissner-Nordstr\"om spacetime~\cite{Zaslavskii:2010aw}.
In this case, we can see the effects of the gravity generated by the colliding objects, 
since Reissner-Nordstr\"om spacetime is more symmetric than Kerr spacetime.  
This will be carried out in the next section. For now, we will ignore the effects of gravity. 

The metric and gauge 1-form of Reissner-Nordstr\"om spacetime is 
\begin{eqnarray}
ds^2 &=& -fdt^2
+
f^{-1}dr^2
+
r^2 (d\theta^2 + \sin^2\theta d\phi^2),
\\
A_a &=& \frac{Q}{r}(dt)_a,
\end{eqnarray}
where the function $f$ is defined by
\begin{eqnarray}
f &=& 1 - \frac{2M}{r} + \frac{Q^2}{r^2},
\end{eqnarray}
$M$ and $Q$ being the mass parameter and the charge, respectively. 
In the case of $M^2>Q^2$, the equation $f=0$ has two positive roots, 
$r=r_\pm:=M\pm\sqrt{M^2-Q^2}$: $r=r_+$ corresponds to the BH horizon, whereas $r=r_-$ 
corresponds to the Cauchy horizon. In the case of $M^2=Q^2$, 
the BH and Cauchy horizons degenerate into one horizon $r=r_\pm=M$. In the case 
of $M^2<Q^2$, there is no real root of $f=0$, i.e., no horizon. 
In this section, we assume $M\geq|Q|$.

The action of a charged test particle subjected to the Lorentz force is given by 
\begin{eqnarray}
S &=& -m\int d\tau - q \int\sum_{\mu=0}^3 A_{\mu} \frac{dx^\mu}{d\tau} d\tau, 
\end{eqnarray}
where $\tau$, $m$ and $q$ are the proper time, inertial mass and charge 
of the test particle, respectively. 
From the minimal action principle, we obtain its equation of motion. 
Without loss of generality, we may assume that the orbit of the particle 
is confined to the equatorial plane $\theta=\pi/2$. We can easily integrate 
the time and azimuthal components of the equation of motion and obtain
\begin{equation}
\frac{dt}{d\tau} = 
\frac{1}{f}\left(
{\cal E}_{\rm c} - \frac{q}{m} A_t
\right)~~~~~~{\rm and}~~~~~~
\frac{d\phi}{d\tau} = \frac{\ell_{\rm c}}{r^2},
\end{equation}
where ${\cal E}_{\rm c}$ and $\ell_{\rm c}$ are integration constants which 
correspond to the specific energy and angular momentum 
of the particle, respectively. 
We assume that ${\cal E}_{\rm c}$ is positive. 
By using the normalization condition of the 4-velocity and the above 
results, we obtain the energy equation
\begin{equation}
\left(\frac{dr}{d\tau}\right)^2 + V = 0,
\label{energyrn}
\end{equation}
where $V$ is the effective potential defined by 
\begin{equation}
V =
-
\left(
{\cal E}_{\rm c} - \frac{q Q}{mr}
\right)^2
+
\left(1 
-
\frac{2 M }{r}
+
\frac{Q^2 }{r^2}
\right)\left(1 + \frac{\ell_{\rm c}^2}{r^2} \right).
\label{potential}
\end{equation}

{}For simplicity, hereafter, we consider a particle radially falling toward 
the black hole, i.e., the case of $\ell_{\rm c}=0$. 
If the background spacetime is extreme $Q = M$ and the charge of 
the particle is $q = {\cal E}_{\rm c} m $, the effective potential becomes
\begin{eqnarray}
V &=&
-
\frac{({\cal E}_{\rm c}^2-1 )(r- M)^2}
{r^2}.
\label{potential-ex}
\end{eqnarray}
{}From the above equation, we can see that this charged test particle 
asymptotically approaches the future degenerate horizon $r = r_\pm =M$, 
i.e., the outward null if ${\cal E}_{\rm c} > 1$.

We consider another particle with an identical inertial mass 
$m$ to the charged particle but with a vanishing charge, 
which is also radially falling toward the black hole. 
Then, let us consider the collision between this neutral particle and the 
charged particle with the effective potential (\ref{potential-ex}). 
We assume that the specific energy 
of the neutral particle is equal to that of the charged particle, ${\cal E}_{\rm c}$.  
We can easily see that, by this assumption, the absolute value of the velocity of the 
neutral particle is larger than that of the charged particle at the same radial coordinate. 
Hence, the charged particle corresponds to the 1st particle, whereas the 
neutral particle corresponds to the 2nd particle in 
Eq.~(\ref{cm}). The square of the 
CM energy at the collision event is obtained as 
\begin{equation}
E_{\rm cm}^2=\frac{2m^2r}{r-M}
\left[1-\frac{M}{r}+{\cal E}_{\rm c}^2-\sqrt{{\cal E}_{\rm c}^2-1}
\sqrt{{\cal E}_{\rm c}^2-\left(1-\frac{M}{r}\right)^2}\right].
\end{equation}
We can easily see that the CM energy diverges in the limit 
$r\rightarrow r_\pm=M$. This is similar to the BSW mechanism. 

It is still difficult to treat the gravity of particles in
Reissner-Nordstr\"om spacetime.  Here, it is worthwhile to note that 
the world line of a radially moving test particle 
is equivalent to the trajectory of an infinitesimally thin 
spherical test shell by virtue of the symmetry 
of Reissner-Nordstr\"om spacetime. 
Since the CM energy at the collision event between two shells 
is the same as that given by Eq.~(\ref{cm}) (see Eq.(\ref{cm-shell}) in 
Sec.\ref{sec:RNshellcollision}), 
an indefinitely large CM energy is realizable also in this case. 
The gravity generated by an 
infinitesimally thin spherical shell can be treated analytically by 
the Israel formalism~\cite{Israel:1966rt}. 
In the next section, we study the effects of the gravity of a thin spherical 
charged shell. 

\section{Infinitesimally thin spherical charged dust shell}\label{sec:RNshell}

\begin{figure}[htbp]
\begin{center}
\includegraphics[width=8cm,clip]{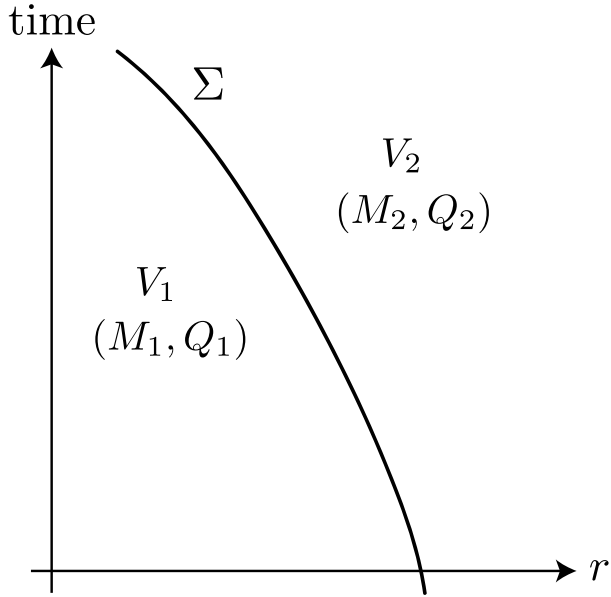}
\end{center}
\caption{
Schematic spacetime diagram of a spherical timelike shell.
The shell divides spacetime into two regions $V_1$ and $V_2$.
}
\label{fig:shell1}
\end{figure}

In this section, we study the gravity generated by a thin spherical shell 
with a non-vanishing charge. 
An infinitesimally thin shell is equivalent to 
a singular timelike hypersurface 
$\Sigma$ which divides spacetime into two regions $V_1$ and $V_2$ 
(see Fig.\ref{fig:shell1}).

The metric tensor $g_{ab}$ should be 
everywhere continuous, even though $\Sigma$ is singular. 
Thus the unit vector $n_a$ normal to $\Sigma$ is uniquely determined, 
and we can introduce 
the intrinsic metric $h_{ab}=g_{ab}-n_a n_a$ on $\Sigma$. 
The extrinsic curvature of $\Sigma$ is defined by 
\begin{equation}
K^{(i)}_{ab}:=-h_a^c h_b^d\nabla^{(i)}_c n_d,
\end{equation} 
which determines how $\Sigma$ is embedded in $V_i$, 
where $\nabla_\alpha^{(i)}$ is the covariant derivative within $V_i$. 
Since the infinitesimally thin shell is a distributional source for the 
Einstein equations, 
$K^{(1)}_{ab}$ and $K^{(2)}_{ab}$ may not be identical to each other.
Following Israel~\cite{Israel:1966rt}, 
we define a tensor field $S_{ab}$ on $\Sigma$ by
\begin{equation}
K^{(2)}_{ab}-K^{(1)}_{ab}=8\pi \left(S_{ab}-\frac{1}{2}h_{ab}
{\rm tr}S\right). 
\label{junction}
\end{equation}
Through the Einstein equation, $S_{ab}$ is identified with 
the surface stress-energy tensor of the shell as 
\begin{equation}
S_{\mu\nu}=\lim_{\varepsilon\rightarrow0}\int_{-\varepsilon}^{+\varepsilon}
T_{\mu\nu}d\lambda,
\end{equation}
where $\lambda$ is the Gaussian normal coordinate in which $\Sigma$ is located at 
$\lambda=0$~\cite{Israel:1966rt}. The above relation implies that the stress-energy tensor of 
the shell can be written in the form
\begin{equation}
T_{\rm shell}^{ab}=S^{ab} \delta(\lambda).
\label{st-tensor-1}
\end{equation}
Further, the Einstein equations lead to
\begin{equation}
D_a S^{ab}=[T_{cd}n^c h^{db}], \label{junction2}
\end{equation}
where $D_a$ is the covariant derivative within $\Sigma$, and the brackets on the 
right hand side of the equation represent the difference 
in a quantity $\Psi$ evaluated on both sides of $\Sigma$: 
$[\Psi]:=\Psi^{(2)}-\Psi^{(1)}$.

We consider a spherically symmetric dust shell whose 
surface stress-energy tensor is given by
\begin{eqnarray}
S^{ab} &=& \sigma u^a u^b,
\end{eqnarray}
where $\sigma$ is the energy density per unit area, which is assumed to be non-negative, 
and $u^a$ can be regarded as the 4-velocity of the shell. 
We assume that this dust shell may have a non-vanishing charge. 
According to the Birkhoff's theorem, the spacetime except on the shell itself 
is Reissner-Nordstr\"om spacetime, and hence 
the metric in the region $V_i$ is given as
\begin{eqnarray}
ds^2 = -f_{(i)}dt_{(i)}^2
+
f^{-1}_{(i)}dr^2
+
r^2(d\theta^2 + {\sin}^2\theta d\phi^2),
\end{eqnarray}
where $f_{(i)}$ is 
\begin{eqnarray}
f_{(i)} =1-\frac{2M_i}{r} + \frac{Q_i^2}{r^2}.
\end{eqnarray}
Here, note that all coordinates except for the time $t$ are common to 
both $V_1$ and $V_2$. 
We assume $M_i \geq |Q_i|$ and denote the 
roots of $f_{(i)}=0$ by $r=r_\pm^{(i)}:=M_i\pm\sqrt{M_i^2-Q_i^2}$. 
In this coordinate system, the components of the 4-velocity $u^a$ are 
\begin{equation}
u^\mu_{(i)}=\left(\frac{dt_{(i)}}{d\tau}, \frac{dr}{d\tau}, 0,0\right),
\end{equation}
where $\tau$ is chosen so that $u^a u_a=-1$.

Using the normalization condition of the 4-velocity, 
Eq.~(\ref{junction2}) leads to 
\begin{eqnarray}
\mu := 4\pi \sigma r^2 = {\rm const.}
\end{eqnarray}
The above equation means that the proper mass $\mu$ of the shell is conserved.

{}From Eq.~(\ref{junction}) and the normalization condition of the 4-velocity,
we obtain the energy equation for the shell as
\begin{equation}
\left(\frac{dr}{d\tau}\right)^2 + V_{\rm shell} = 0,
\label{energyrnshell}
\end{equation}
where the effective potential $V_{\rm shell}$ is written in the form
\begin{equation}
V_{\rm shell} =
-
\left(
{\cal E} - \frac{q \langle Q \rangle}{\mu r}
\right)^2
+
1 
-
\frac{2\langle M \rangle }{r}
+
\frac{\langle Q \rangle^2}{r^2}
-
\left(\frac{\mu}{2r}\right)^2
+
\left(\frac{q}{2r}\right)^2,
\label{potential-shell}
\end{equation}
where
\begin{equation}
{\cal E} := \frac{M_2 - M_1}{\mu},~~ 
\langle M \rangle := \frac{M_2 + M_1}{2},~~ 
\langle Q \rangle := \frac{Q_2 + Q_1}{2}~~{\rm and}~~q := Q_2 - Q_1.
\end{equation}
Here note that $\mu {\cal E}=M_2-M_1$ is equal to the Misner-Sharp energy 
concentrated on the shell~\cite{Misner:1964je,Hayward:1994bu}. Thus, we may call ${\cal E}$ the specific 
energy of the shell, and we assume that it is positive. 
We can see that the effective potential (\ref{potential-shell}) 
has almost the same form as that for a charged test particle (\ref{potential}) 
with $\ell_{\rm c} = 0$, or equivalently, that for a spherical charged test shell. 
The differences between Eqs.~(\ref{potential}) with $\ell_{\rm c}=0$ 
and (\ref{potential-shell}) are 
regarded as the self-gravity and self-electric interaction terms.

Let us investigate whether the charged dust shell can 
asymptotically approach an outward null hypersurface as 
in the case of the charged test shell. 
The outside of the shell is the region $V_2$, 
and the BH horizon in this region is $r = r_+^{(2)}:=M_2 + \sqrt{M_2^2 - Q_2^2}$ 
which is the outward null hypersurface. 
Hence, we search for the condition which guarantees $r=r_+^{(2)}$ to be 
the asymptote of the singular hypersurface $\Sigma$. 
This task is equivalent to solving $V(r_+^{(2)}) = 0$ and $dV(r)/dr|_{r=r_+^{(2)}} = 0$ 
in terms of the parameters, $Q_2$, $M_2$, $\mu$, $q$ and ${\cal E}$. We have
\begin{eqnarray}
Q_2 &=& M_2, \label{condition1}
\\
q^2 - 2 M_2 q + 2 M_2 {\cal E} \mu - \mu^2 &=&0. \label{condition2}
\end{eqnarray}
Condition (\ref{condition1}) implies that region $V_2$ is an extreme 
Reissner-Nordstr\"om spacetime. If the above conditions hold, 
the effective potential becomes
\begin{eqnarray}
V_{\rm shell}&=& - \frac{({\cal E}^2 - 1)(r- M_2)^2}{r^2}.
\end{eqnarray}
Namely, if this charged dust shell is contracting, it 
asymptotically approaches the outward 
null hypersurface $r =r_\pm^{(2)}= M_2$ as long as ${\cal E} > 1$. 
Hence, it is likely that a BSW-type mechanism can occur in this system. 
It is worthwhile to note 
that this result is correct even if there is no central black hole, i.e., 
$M_1 = Q_1 = 0$.

\section{effect of gravity generated by colliding shells in BSW process}\label{sec:RNshellcollision} 
\begin{figure}[htbp]
\begin{center}
\includegraphics[width=8cm,clip]{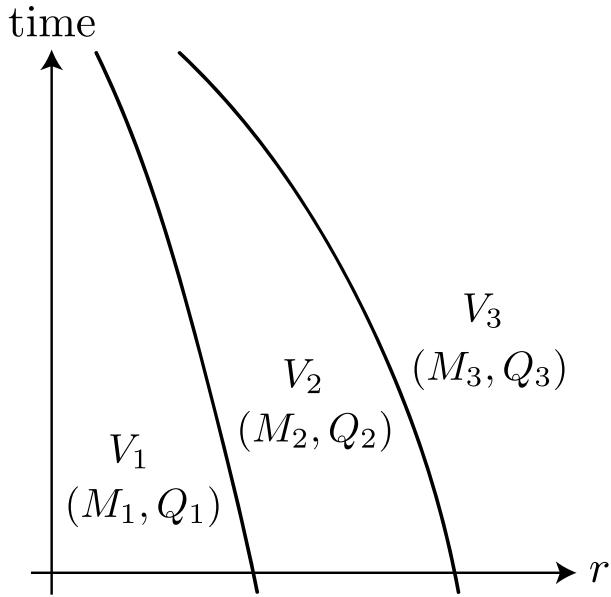}
\end{center}
\caption{
Schematic spacetime diagram of two spherical timelike shells.
The two shells divide the spacetime into three regions $V_1$, $V_2$ and $V_3$.
}
\label{fig:shell2}
\end{figure}
In this section, we consider collisions of two spherical dust shells and discuss
the CM energy at the collision event. 
These two shells divide the spacetime into three regions, $V_1$, 
$V_2$ and $V_3$, before the collision (see Fig.\ref{fig:shell2}).
We assume that the inner shell is the same as that considered in the 
preceding section, whose parameters satisfy 
conditions (\ref{condition1}) and (\ref{condition2}). 
We also assume that, as in Sec.~\ref{sec:RN},  
the outer shell is composed of neutral dust and 
has the same specific energy ${\cal E}$ as the inner charged shell.  
Because the outer shell is neutral, we have $Q_2=Q_3$. 
Further, both shells are assumed to have an identical proper mass, 
$\mu_{\rm out} = \mu_{\rm in} = \mu$.

The total stress-energy tensor of the shells is written in the form
\begin{equation}
T_\Sigma^{ab}=
\sigma_{\rm in}u_{\rm in}^a u_{\rm in}^b \delta(\lambda_{\rm in})
+\sigma_{\rm out}u_{\rm out}^a u_{\rm out}^b \delta(\lambda_{\rm out})
\label{st-tensor}
\end{equation}
where $\lambda_{\rm in}$ and $\lambda_{\rm out}$ are the Gaussian normal coordinates 
of the inner and outer shells, respectively. 
Since the proper masses $\mu$ of these shells are identical to each other, 
$\sigma_{\rm in}$ is equal to $\sigma_{\rm out}$ at the collision event. 

Here, we introduce the ``center of mass frame'' at the collision event of 
the shells. It is composed of the 
orthonormal basis $(\bar{u}^a, \bar{n}^a,e_{(\theta)}^a,e_{(\phi)}^a)$ 
which satisfies the following condition
\begin{equation}
T_\Sigma^{ab}\bar{u}_a \bar{n}_b=0=
T_\Sigma^{ab}\bar{u}_a e_{(\theta)b}=
T_\Sigma^{ab}\bar{u}_a e_{(\phi)b}.
\end{equation}
The above condition means that the spatial components of the 
energy flux vanish in this frame. Then, the energy of the colliding shells in
this frame, which is also called the CM energy $E_{\rm cm}$, is given by
\begin{equation}
E_{\rm cm}=\int T_\Sigma^{ab}\bar{u}_a \bar{u}_b r^2 \sin\theta 
d\bar{\lambda} d\theta d\phi
=\sqrt{2}\mu\sqrt{1-g_{ab}u_{\rm in}^a u_{\rm out}^b},
\label{cm-shell}
\end{equation}
where $\bar{\lambda}$ is the proper length in the direction of $\bar{n}^a$. 
As expected, the CM energy of the shells takes the same form as that of the 
particles. Details of this derivation are given in the Appendix. 

As shown in the preceding section, the outward null hypersurface 
$r=r_\pm^{(2)}=M_2$ is the asymptote of the inner shell. Thus, as in the case of 
the test shells, the closer the relative velocity between the inner and outer shells 
approaches to the speed of light, the closer the collision event 
approaches to the BH horizon in $V_2$, i.e., $r=r_\pm^{(2)}$. As a result, the 
CM energy at the collision event can be indefinitely large, even if 
the gravity of the colliding shells is taken into account. However, 
we should note that if the two shells collide inside the BH horizon in $V_3$, i.e., 
$r \leq r_+^{(3)} = M_3 + \sqrt{M_3^2 - Q_3^2}$, 
distant observers like us could not see the collision of these shells. 
We should also note that $r_+^{(3)}$ is larger than $r_\pm^{(2)}$ by virtue 
of the gravity generated by the outer shell. 
Thus, the observable CM energy is less than that of the collision 
at $r=r_+^{(3)}$ 
given in accordance with Eq.~(\ref{cm-shell}) as
\begin{eqnarray}
E_{\rm cm}(r=r_+^{(3)}) =
\sqrt{2}\mu\sqrt{1-
\frac{1}{\sqrt{\mu}}
\frac{Y}{2 \sqrt{{\cal E} (3 {\cal E}
   \mu+2 M_1)}
+
2 {\cal E}\sqrt{\mu}}},
\end{eqnarray}
where the function $Y$ is given by
\begin{eqnarray}
Y &=& 
\sqrt{
{\cal E}^2-1
}
\sqrt{
2 ({\cal E}
   \mu+M_1)+2 ({\cal E}-1)
   \left(\sqrt{{\cal E} \mu (3
   {\cal E} \mu+2 M_1)}+2
   {\cal E}
   \mu+M_1\right)+\mu
}
\notag\\ && \times
\sqrt{
-2 ({\cal E}
   \mu+M_1)+2 ({\cal E}+1)
   \left(\sqrt{{\cal E} \mu (3
   {\cal E} \mu+2 M_1)}+2
   {\cal E}
   \mu+M_1\right)+\mu
}
\notag\\ &&
-
{\cal E} \left(2 {\cal E}
   \left(\sqrt{{\cal E} \mu (3
   {\cal E} \mu+2 M_1)}+2
   {\cal E}
   \mu+M_1\right)+\mu
   \right).
\end{eqnarray}
We can see from the above result that the CM energy at the collision event between 
these shells has an upper limit in the observable domain 
$r>r_+^{(3)}$, and this limit is determined by the mass $M_1$ of the 
``central black hole'' and the proper mass of the two shells. 
Here, we should note that in order that the outer shell overtakes the inner shell, 
$(u^r_{\rm out})^2-(u^r_{\rm in})^2$ should be positive at $r=r^{(3)}_+$. 
We can see that this condition holds from
\begin{eqnarray}
\left[(u^r_{\rm out})^2-(u^r_{\rm in})^2\right]_{r=r^{(3)}_+}
&=& 
\left[2 {\cal E} ({\cal E}
   \mu+M_1)+\mu \right]
\notag \\ && \times
\left[6 {\cal E}^2 \mu+4 {\cal E}
   \sqrt{{\cal E} \mu (3
   {\cal E} \mu+2 M_1)}+2
   {\cal E} M_1+\mu\right]
\notag \\ && \times
\left[4 (\sqrt{{\cal E} \mu (3
   {\cal E} \mu+2 M_1)}+2
   {\cal E}
   \mu+M_1)^2\right]^{-1} 
\notag \\ &>& 0.
\end{eqnarray}

In the case that the mass $M_1$ of the central black hole is much 
larger than the proper masses of the shells 
$\mu$, the observable CM energy becomes
\begin{eqnarray}
E_{\rm cm} \simeq
2^{1/4}
{\cal E}^{1/4}
\sqrt{{\cal E} - \sqrt{{\cal E}^2 - 1}} 
M_1^{1/4}\mu^{3/4}. 
\label{large-BHmass}
\end{eqnarray}
We can see from the above equation that, also in this case, 
the observable CM energy is not indefinitely large. 
Thus, when estimating the size of the observable CM energy, the gravity caused by the colliding objects 
must not be ignored, even if their initial energy is very small.

\section{SUMMARY AND DISCUSSION}\label{sec:summary}
In this paper, we studied a collision of two 
spherical dust shells in order to determine the effects of the gravity generated by the 
colliding objects in the BSW process. 
We have shown that a contracting charged dust 
shell can asymptotically approach the outward null hypersurface, even if its gravity is taken into account. 
If such a shell collides with another contracting shell, the relative velocity 
between them can be arbitrarily close to the speed of light. 
However, we found that the CM energy of two colliding shells has an upper limit 
in the observable domain, since the event horizon moves 
outward due to the gravity of the outer shell. 
Our results suggest that two particles can not collide 
with an arbitrarily high energy in the center of mass frame 
in the observable domain, if we take into account the effects of the gravity of the colliding particles.

Each dust shell can be regarded as an aggregation of many particles. 
Thus, it is worthwhile to study how a large CM energy of the constituent particles 
of the shells can be achieved by a collision of two shells. 
We assume that the mass of the central 
black hole $M_1$ is much larger than that of a shell $\mu$, 
and the constituent particles have an identical rest mass $m$. 
Thus, the number $N$ of particles included in a shell 
is given by $N=\mu/m$. The energy $E$ of a particle in the center of mass frame is 
given by $E=E_{\rm cm}/2N$. 
In this section, we denote the speed of light and the gravitational constant by 
$c$ and $G$, respectively. 
Then, from Eq.~(\ref{large-BHmass}), we have
\begin{equation}
\frac{E}{mc^2}=\frac{E_{\rm cm}}{2\mu c^2}\simeq \alpha\left(\frac{M_1}{\mu}\right)^{1/4},
\end{equation}
where 
\begin{equation}
\alpha=
\frac{1}{2^{3/4}}{\cal E}^{1/4}\sqrt{{\cal E} - \sqrt{{\cal E}^2 -1}}.
\end{equation}
We can easily see that $\alpha$ is a monotonically decreasing function of ${\cal E}$ 
for ${\cal E}\geq1$ and $\alpha=2^{-3/4}\simeq 0.59$ at ${\cal E}=1$. 
We again note that we are interested in the case of ${\cal E}>1$. 
Here, we introduce a parameter defined by
\begin{equation}
\beta=\frac{E}{m_{\rm pl}c^2},
\end{equation}
where $m_{\rm pl}$ is the Planck mass ($2.18\times10^{-5}$ g).  
When the shells collide with each other near the horizon $r=GM_1/c^2$, 
the mean separation between the constituent particles of the shells is given by
\begin{equation}
\ell \simeq \sqrt{\frac{4\pi (GM_1/c^2)^2}{2N}}
\simeq 1.2
\left(\frac{\beta}{\alpha}\right)^{2}
\left(\frac{10^{-9}m_{\rm pl}}{m}\right)^{{3/2}}
\left(\frac{M_1}{M_\odot}\right)^{1/2}{\rm cm},
\end{equation}
where $M_\odot$
is the solar mass, 
respectively. The above equation implies that the mean separation is 
much larger than the Compton wavelength if 
the mass $M_1$ of the central black hole is equal to $M_\odot$, 
if the mass $m$ of a constituent particle is equal to $10^{-9}$ times the 
Planck mass $m_{\rm pl}$ and if the mean CM energy $E$ of a constituent particle 
is equal to or larger than the Planck energy, i.e., $\beta \geq 1$. 
This macroscopic separation implies that the dust approximation 
can be valid, even if $E$ is super-Planckian. 
Further, since the mean separation $\ell$ can be much smaller 
than $r=GM_1/c^2\simeq 1.5\times 10^5(M_1/M_\odot)$ cm, 
the continuum approximation can also be valid. 
Thus, although the mass of each constituent particle has to be rather large, 
we may say that the CM energy of the constituent particle can be super-Planckian 
within the dust shell approximation, even if it cannot be indefinitely large.

The CM energy of a particle, $E$, is rewritten from Eq. (43) as,
\begin{eqnarray}
\beta=\frac{E}{m_{\rm pl}c^2}=1.4\times 10^{-5} N^{-1/4} \alpha \left(\frac{m}{m_{\rm proton}}\right)^{3/4}
\left(\frac{M_1}{M_\odot}\right)^{1/4},
\end{eqnarray}
where $m_{\rm proton}$ is the proton mass.
Thus, within the dust shell approximation, since $N$ is much larger than unity,
the CM energy will be smaller than the Planck energy significantly
if the mass of the constituent particles is equal to the proton mass.
If we were allowed to extrapolate this equation to $N\sim 1$,
we would obtain $E\simeq10^{-5}m_{\rm pl}$ in the case $m=m_{\rm proton}$, $M_1=10M_\odot$ and $\alpha=0.59$.
Of course, since such an extrapolation might not be very accurate,
much more detailed investigation will be necessary to evaluate 
an accurate value of the maximum CM energy.

\section*{Acknowledgments}
We would like to thank Tomohiro Harada, Hideki Ishihara, 
Umpei Miyamoto and Takahiro Tanaka for useful discussions.
H.T.'s work was supported in part
by a Monbu Kagakusho Grant-in-aid
for Scientific Research of Japan (No. 20540271).

\appendix
\section{Center of mass frame for spherical shell}
\label{section:cm-frame}

The components of the orthonormal basis of the ``center of mass frame'' 
for two spherical shells considered in 
Sec.\ref{sec:RNshellcollision} are denoted as
\begin{eqnarray}
\bar{u}^a&=&
\left(\frac{dt}{d\bar{\tau}},\frac{dr}{d\bar{\tau}},0,0\right), \\
\bar{n}^a&=&
\left(\frac{dt}{d\bar{\lambda}},\frac{dr}{d\bar{\lambda}},0,0\right), \\
e_{(\theta)}^a
&=&\left(0,0,\frac{1}{r},0\right), \\
e_{(\phi)}^a
&=&\left(0,0,0,\frac{1}{r\sin\theta}\right). 
\end{eqnarray}
By using these basis vectors, the 4-velocities of the two shells 
are expressed in the form
\begin{eqnarray}
u_{\rm in}^a&=&\alpha_{\rm in}\bar{u}^a+\beta_{\rm in}\bar{n}^a, \label{u-in}\\
u_{\rm out}^a&=&\alpha_{\rm out}\bar{u}^a+\beta_{\rm out}\bar{n}^a, \label{u-out}
\end{eqnarray}
where $\alpha_{\rm in}$ and $\alpha_{\rm out}$ are assumed to be positive. 
The normalization of the 4-velocities leads to
\begin{equation}
\alpha_{\rm in}^2-\beta_{\rm in}^2=1~~~~~~{\rm and}~~~~~~
\alpha_{\rm out}^2-\beta_{\rm out}^2=1.
\label{normalization}
\end{equation}
By the normalization and orthogonal conditions, 
the normal vectors to the two shells are given by
\begin{eqnarray}
n_{\rm in}^a&=&\beta_{\rm in}\bar{u}^a+\alpha_{\rm in}\bar{n}^a, \label{normal-in}\\
n_{\rm out}^a&=&\beta_{\rm out}\bar{u}^a+\alpha_{\rm out}\bar{n}^a. \label{normal-out}
\end{eqnarray}

The stress-energy tensor of the shells (\ref{st-tensor}) at the 
collision event is expressed 
by using the basis of the CM frame as
\begin{eqnarray}
T_\Sigma^{ab}&=&\sigma
\Biggl[
\left\{
\alpha_{\rm in}^2
\left(\frac{\partial\lambda_{\rm in}}{\partial\bar{\lambda}}\right)_{\bar\tau}^{-1}
+\alpha_{\rm out}^2
\left(\frac{\partial\lambda_{\rm out}}{\partial\bar{\lambda}}\right)_{\bar\tau}^{-1}
\right\}\bar{u}^a \bar{u}^b \nonumber \\
&+&\left\{
\alpha_{\rm in}\beta_{\rm in}
\left(\frac{\partial\lambda_{\rm in}}{\partial\bar{\lambda}}\right)_{\bar\tau}^{-1}
+\alpha_{\rm out}\beta_{\rm out}
\left(\frac{\partial\lambda_{\rm out}}{\partial\bar{\lambda}}\right)_{\bar\tau}^{-1}
\right\}(\bar{u}^{a} \bar{n}^{b}+\bar{u}^{b} \bar{n}^{a}) \nonumber \\
&+&\left\{
\beta_{\rm in}^2
\left(\frac{\partial\lambda_{\rm in}}{\partial\bar{\lambda}}\right)_{\bar\tau}^{-1}
+\beta_{\rm out}^2
\left(\frac{\partial\lambda_{\rm out}}{\partial\bar{\lambda}}\right)_{\bar\tau}^{-1}
\right\}\bar{n}^a \bar{n}^b
\Biggr]\delta(\bar{\lambda}),
\end{eqnarray}
where we have assumed that $\sigma_{\rm in}=\sigma_{\rm out}=\sigma$ at the 
collision event.

The condition $T^{ab}_\Sigma \bar{u}_a \bar{n}_b=0$ for the CM frame leads to
\begin{equation}
\alpha_{\rm in}\beta_{\rm in}
\left(\frac{\partial\lambda_{\rm in}}{\partial\bar{\lambda}}\right)_{\bar\tau}^{-1}
+\alpha_{\rm out}\beta_{\rm out}
\left(\frac{\partial\lambda_{\rm out}}{\partial\bar{\lambda}}\right)_{\bar\tau}^{-1}
=0.
\label{cm-condition}
\end{equation}
From the definition of the orthonormal frame, we have 
\begin{equation}
n_{\rm in}^a\frac{\partial \lambda_{\rm in}}{\partial x^a}=1~~~~~{\rm and}~~~~~ 
u_{\rm in}^a\frac{\partial\lambda_{\rm in}}{\partial x^a}=0,
\end{equation}
and thus we have 
\begin{equation}
g^{ab}\frac{\partial \lambda_{\rm in}}{\partial x^b} =n_{\rm in}^a.
\end{equation} 
By using the above relation, Eqs.~(\ref{normal-in}) and (\ref{normal-out}) lead to
\begin{equation}
\left(\frac{\partial \lambda_{\rm in}}{\partial\bar{\lambda}}\right)_{\bar{\tau}}
=n_{\rm in}^a \bar{n}_a=\alpha_{\rm in}.
\end{equation}
In the same manner as above, we have
\begin{equation}
\left(\frac{\partial \lambda_{\rm out}}{\partial\bar{\lambda}}\right)_{\bar{\tau}}
=\alpha_{\rm out}.
\end{equation} 
Thus, Eq.~(\ref{cm-condition}) becomes
\begin{equation}
\beta_{\rm in}+\beta_{\rm out}=0.
\label{cm-condition2}
\end{equation}
By Eq.~(\ref{normalization}), the above equation leads to
\begin{equation}
\alpha_{\rm in}^2=\alpha_{\rm out}^2. 
\end{equation}
Thus, we have $\alpha_{\rm in}=\alpha_{\rm out}=:\alpha$. 
Further, by Eqs.~(\ref{u-in}) and (\ref{u-out}) and one of the 
orthonormal conditions for $u^a$ and $n^a$, we obtain
\begin{equation}
\alpha=\sqrt{\frac{1}{2}\left(1-g_{ab}u_{\rm in}^a u_{\rm out}^b\right)}.
\end{equation}
Hence we have
\begin{equation}
T_\Sigma^{ab} \bar{u}_a \bar{u}_b=\sqrt{2}\sigma
\sqrt{1-g_{ab}u_{\rm in}^a u_{\rm out}^b}\delta(\bar{\lambda}).
\end{equation}
The integration of the above quantity over 
$(\bar{\lambda},\theta,\phi)$ gives ``the CM energy'' of the two shells.

\end{document}